# Clinical Feasibility of Smartphone-based EEG in Kenya

William Lehn-Schiøler[1–2], Nomin Enkhtsetseg[1], Anton Mosquera Storgaard[1], Magnus Guldberg Pedersen[1], Dylan Rice[1], George Wambugu[1], Nshimiyimana Jules Fidele[3], Melita Cacic Hribljan[4–5], Anca Adriana Arbune[6], Sidsel Armand Larsen[7–8], Sándor Beniczky[7–9]

1 BrainCapture, Kongens Lyngby, Denmark
2 Department of Health Technology, Technical University of Denmark, Kongens Lyngby, Denmark
3 Kenyatta University Teaching, Referral and Research Hospital, Nairobi, Kenya
4 Department of Clinical Neurophysiology, Copenhagen University Hospital Rigshospitalet, Copenhagen, Denmark
5 Department of Pediatrics, Children's Hospital Srebrnjak, Zagreb, Croatia
6 Department of Stress Prophylaxis and Research, "Prof. Dr. Alexandru Obregia" Psychiatry Hospital, Bucharest, Romania
7 Danish Epilepsy Center, Filadelfia, Denmark
8 Copenhagen University Hospital Rigshospitalet, Copenhagen, Denmark
9 Department of Clinical Medicine, University of Copenhagen, Copenhagen, Denmark

**Corresponding author:**
Sándor Beniczky, MD, PhD
Department of Clinical Medicine, University of Copenhagen
Copenhagen University Hospital - Rigshospitalet
Blegdamsvej 9
2100 Copenhagen Ø, Denmark
Email: sbz@filadelfia.dk

**Abstract:**

Purpose: Access to electroencephalography (EEG) remains limited across low- and middle-income countries (LMICs) due to cost, infrastructure requirements, and a shortage of trained staff. This study evaluated the feasibility and clinical utility of a smartphone-based EEG system in a real-world setting.

Methods: We conducted a multicenter observational study (November 2023 to April 2026) across 29 clinical sites in Kenya. A smartphone-based 27-lead EEG system enabled trained healthcare workers to acquire standardized recordings with remote expert interpretation.

Results: 3,036 EEG sessions were performed. Male patients constituted 57.8% of the cohort, with representation across pediatric and adult populations. The most common referral indication was seizures or convulsions (68.5%). Overall, 2,915 (96%) recordings were interpretable, while 121 (4%) were uninterpretable, primarily due to high electrode impedance and insufficient recording duration. Uninterpretable recordings were significantly shorter than interpretable recordings (mean 18.5 vs. 33.8 minutes; median 15.1 vs. 31.6 minutes; $p < 0.0001$). Mean turnaround time for interpretation was 107 minutes.

Among interpretable recordings, 917 (30.2%) were abnormal, including 701 (76.4%) with epileptiform abnormalities, 215 (23.4%) with non-epileptiform findings, and 1 (<0.1%) indeterminate finding. Epileptiform abnormalities were highest in children aged 4–9 years (33.1%) and less frequent in adults (14–21%). Non-epileptiform abnormalities were more common in patients aged ≥60 years (19.2%) compared to younger age groups (3–9%).

Conclusion: Large-scale, point-of-care EEG acquisition by non-specialist operators in a resource-limited setting is feasible. Expansion of smartphone-based EEG systems may improve equitable access to neurological diagnosis and care in LMICs.

## Introduction

Epilepsy is one of the most common neurological disorders globally, and a major contributor to neurological disability worldwide [1–3], affecting individuals of all ages, with over 80% of the burden occurring in low- and middle-income countries (LMICs) [4]. Despite this, access to essential diagnostic tools such as electroencephalography (EEG) remains limited, especially in LMICs, where infrastructure, trained personnel, and centralized services restrict availability [5]. As a result, many individuals with epilepsy remain undiagnosed or misdiagnosed, contributing to gaps in treatment and preventable morbidity [6].

Conventional EEG systems require advanced equipment, dedicated facilities, and technical expertise, posing significant challenges to implementation in many LMIC clinical settings [7]. These barriers often concentrate neurological services in urban tertiary care centers, leaving large rural populations without access to diagnostic evaluation [5,7–8].

Kenya, a lower-middle-income country [9] with a high burden of neurological disease and limited access to neurodiagnostic services, provides a relevant setting to evaluate scalable approaches to expanding EEG access [10–12]. Elevated epilepsy burden in such settings demonstrates both increased exposure to preventable neurological risk factors (e.g., traumatic brain injury, severe malaria, perinatal injury, and infections) [13–14] and systemic barriers to care, including limited diagnostic infrastructure, workforce shortages, financial and transportation constraints, and cultural stigma [15–16].

Recent advances in portable and mobile neurodiagnostic technologies offer potential solutions to expand EEG access beyond traditional clinical environments. While prior studies have demonstrated the feasibility and acceptability of portable EEG technologies in LMIC and community-based settings [17–19], evidence remains limited regarding large-scale clinical implementation and diagnostic yield in routine care settings. These approaches align with priorities outlined in the World Health Organization's (WHO) Intersectoral Global Action Plan on Epilepsy and Other Neurological Disorders (IGAP) [3], which emphasizes improving diagnostic capacity and reducing inequities in access to neurological care.

In this study, we report data from more than 3,000 routine clinical smartphone-based EEGs performed across 29 sites in Kenya. Using this data, we (1) evaluate the feasibility of supplementing routine care using smartphone-based EEG recordings across diverse clinical settings, (2) characterize the prevalence of epileptiform findings; (3) examine age-related diagnostic yield; and (4) explore clinical indications and corresponding abnormal findings. These findings provide large-scale, real-world foundational evidence on the implementation of smartphone-based EEG in outpatient settings in a lower-middle-income country.

## Materials and Methods

*Study Design*

This was a retrospective, multicenter observational study evaluating the clinical utility and feasibility of a smartphone-based EEG system. Data were collected from November 2023 through the study data cutoff date of April 2026 (approximately 29 months).

*Setting*

The study collected data across 29 healthcare facilities in Kenya, ranging from urban clinical imaging centers to smaller rural facilities with previously limited or non-existent neurodiagnostic infrastructure (Figure 1). All EEGs performed during this timeframe were part of the study's analysis.

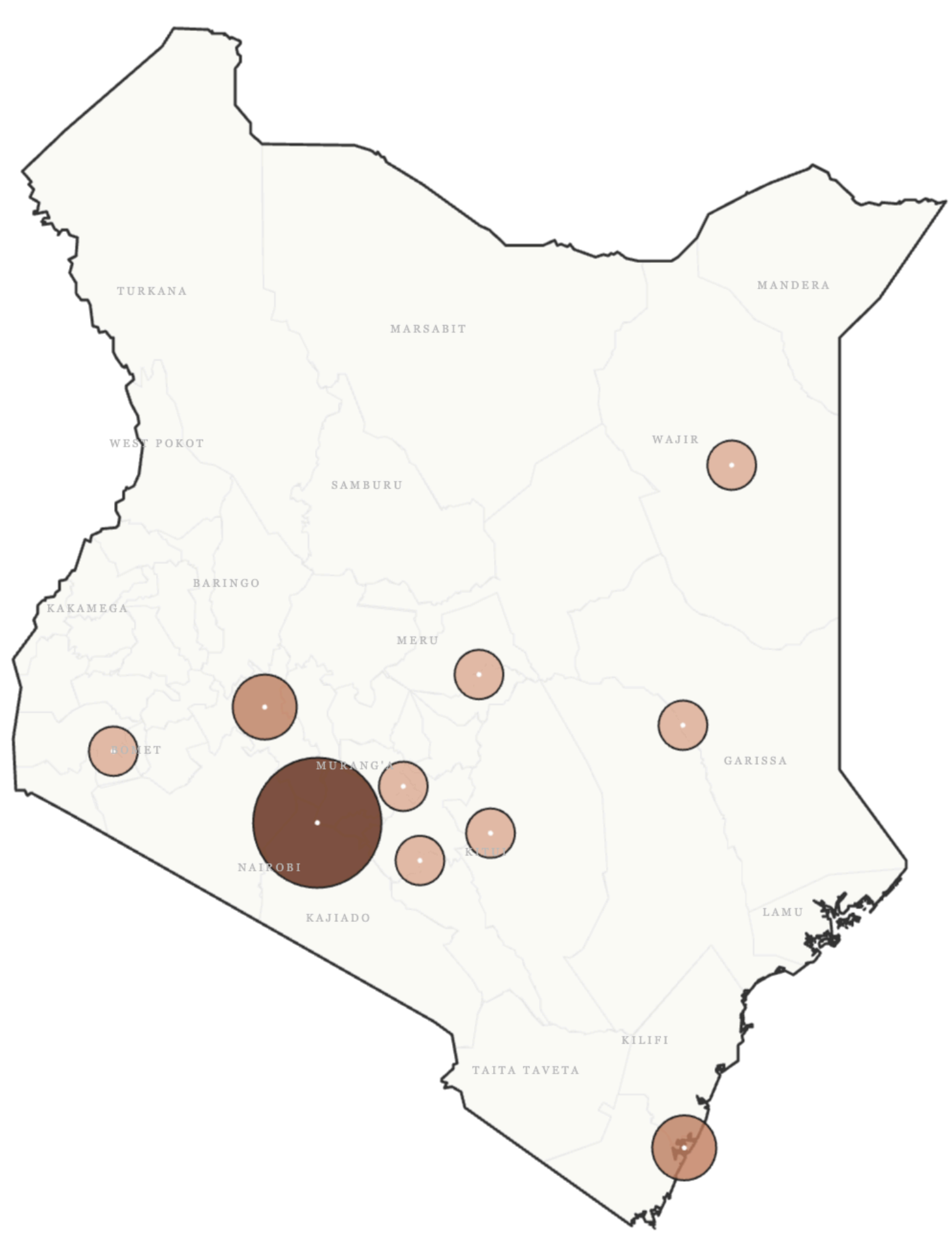

Figure 1. Map of Kenya demonstrating the distribution of EEG recordings across various regions *(2-column)*

*Participants*

Patients were deemed eligible if they underwent a smartphone-based EEG recording at one of the 29 participating sites during the study period and provided consent. No additional exclusion criteria were applied to reflect real-world clinical use. EEG referrals were made by the treating clinician based on routine clinical indications, and no follow-up procedures were performed.

*Equipment and Procedure*

BrainCapture (https://www.braincapture.dk/), a Danish medical technology company, designed the BC-1 system as a smartphone-based EEG device intended to be used in resource-limited environments [20]. The device consists of a smartphone-connected amplifier and a stretchable cap with pre-positioned electrodes designed to align with the international 10-20 system (Figure 2) [20]. The smartphone-based EEG aims to reduce infrastructure requirements, including the need for dedicated recording facilities and expert personnel, while enabling data acquisition in decentralized environments.

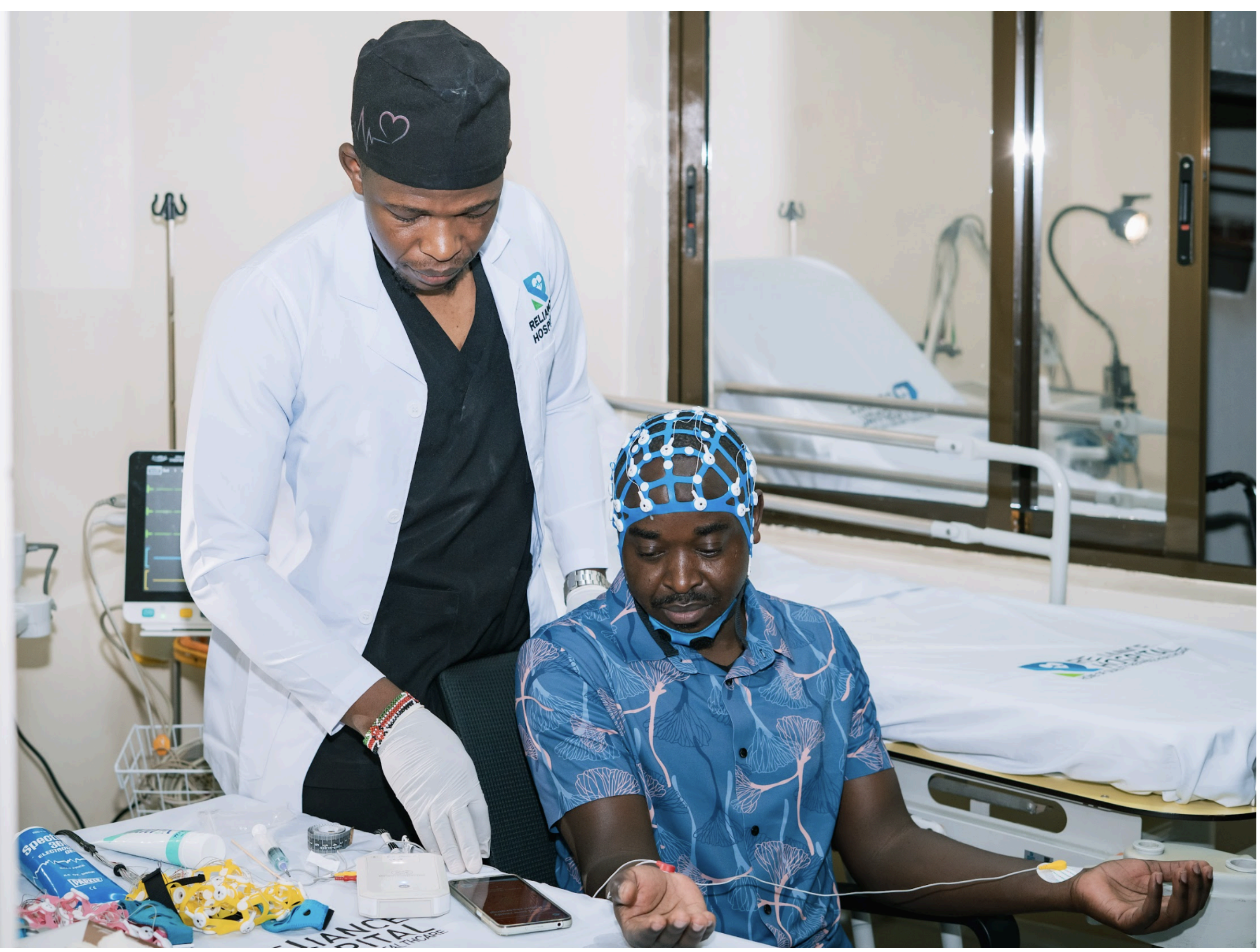

Figure 2. Implementation of smartphone-based EEG (BrainCapture BC-1) in a clinical setting. *(1.5-column)*

EEG recordings were performed by local healthcare staff (e.g., nurses and clinical officers) who received standardized minimal training in skin preparation, cap placement, and the use of the BrainCapture mobile application. This brief training included a short theoretical introduction to EEG and epilepsy, followed by a supervised hands-on practice with two EEG recordings prior to independent data acquisition. Each session followed a routine clinical protocol lasting approximately 20–30 minutes. The system utilized the smartphone's cellular or Wi-Fi connectivity to upload EEG data to a secure cloud-based platform. Real-time impedance checks were integrated into the mobile app to guide the technician in ensuring signal quality before and during the recording.

*Data Interpretation*
Recorded EEGs were reviewed by a certified EEG technologist and validated by a neurophysiologist. Interpretations were categorized according to standard clinical neurophysiology guidelines [21]. Recordings were flagged by the reader as "uninterpretable" if excessive artifacts (e.g., movement, muscle, or technical interference) or high impedance precluded a reliable clinical conclusion. Interpretable EEGs were classified as “Normal” or “Abnormal” and the abnormalities were further sub-classified into “Epileptiform” or “Non-epileptiform”.

*Clinical and Demographic Data*
For each EEG performed, basic demographic and clinical information were captured through a short survey conducted via the smartphone application at the time of the recording, including patient demographic variables, clinician-reported indication for EEG, and medication history.

*Variables*
The primary outcomes of the study were EEG feasibility and clinical utility. Feasibility was assessed by EEG interpretability rates (i.e., proportion of recordings that were successfully completed and deemed interpretable) and successful deployment across study sites. Clinical utility was defined as the ability of smartphone-based EEG to provide actionable diagnostic information through demonstrating abnormal findings, including epileptiform activity, that could inform clinical decision-making. Secondary measures included turnaround time from recording completion to interpretation, recording duration, and the distribution of EEG findings across demographic and clinical subgroups.

*Sample Size*
The sample size was determined by the total number of EEGs performed by the target censoring date of April 6th, 2026.

*Statistical Analysis*
Descriptive statistics were used to characterize population characteristics, EEG findings, and clinical indications. Continuous variables were reported using means and medians with categorical variables as proportions. Missing data was assessed for each variable and reported where appropriate. Primary analyses involved the prevalence of EEG abnormalities, including epileptiform and non-epileptiform findings, as well as exploring their distribution across age and clinical indication groups. Comparative analysis of interpretable and uninterpretable recording characteristics was conducted with statistical significance defined as a two-sided p-value $<0.05$.

*Ethics*
Data were de-identified at the source to ensure patient confidentiality in accordance with local data protection regulations.

## Results

*Study Cohort and Feasibility*
A total of 3,036 EEG sessions were performed across 29 sites in Kenya between November 2023 and April 2026. EEG recordings were more frequently performed on male patients (n=1,756, 57.8%) compared to female patients (n=1,280, 42.2%).

Among the 3,036 EEG recordings included in the final analysis, 2,915 (96%) were judged clinically interpretable by remote interpreters, while 121 (4%) were uninterpretable (Figure 3). To characterize factors associated with uninterpretable recordings, impedance and recording duration were compared between the 121 uninterpretable EEGs and a matched sample of interpretable recordings. Uninterpretable recordings were significantly shorter than interpretable ones (mean 18.5 vs. 33.8 minutes, median 15.1 vs. 31.6 minutes; $p < 0.0001$), with only 45% exceeding 20 minutes compared to 89% of interpretable recordings. Uninterpretable recordings also demonstrated significantly fewer channels with adequate electrode-skin contact across all impedance thresholds ($p < 0.0001$), indicating more widespread electrode failure. Together, these findings suggest that the 4% uninterpretable recordings were primarily driven by two compounding factors: insufficient recording duration and suboptimal electrode connectivity, both consistent with the operational challenges of community-based EEG acquisition performed by trained lay operators in a low-resource field setting.

Among recordings with available turnaround time data, most interpretations were completed within 1-2 hours, with a mean interpretation time of 107 minutes and a median of 102 minutes.

Demographic and clinical characteristics of the study population are summarized in Table 1.

Table 1. Demographic and clinical characteristics of the study cohort

| **Characteristic** | **n (%)** |
|---|---|
| Total EEG recordings | 3,036 |
| Interpretable recordings | 2,915 |
| **Sex** | |
| Male | 1,756 (57.8%) |
| Female | 1,280 (42.2%) |
| **Age group** | |
| 0–3 years | 470 (15.5%) |
| 4–9 years | 513 (16.9%) |
| 10–19 years | 708 (23.3%) |
| 20–29 years | 422 (13.9%) |
| 30–39 years | 283 (9.3%) |
| 40–49 years | 155 (5.1%) |
| 50–59 years | 116 (3.8%) |
| 60+ years | 234 (7.7%) |
| Unknown age | 135 (4.4%) |
| **Primary clinical indication** | |
| Seizures/convulsions | 2,107 (69.4%) |
| Loss of consciousness | 129 (4.3%) |
| Headache | 55 (1.8%) |
| Altered consciousness | 45 (1.5%) |
| Behavioral/psychiatric | 39 (1.3%) |
| Non-Symptom/Follow-up | 27 (0.9%) |

| Structural injury | 24 (0.8%) |
| --- | --- |
| Spells | 12 (0.4%) |
| Sensory/numbness | 8 (0.3%) |
| Mood/depression | 7 (0.2%) |
| Drop (attacks/spells) | 6 (0.2%) |
| Stroke | 6 (0.2%) |
| Staring | 3 (0.1%) |
| Anxiety | 2 (0.1%) |
| Pain | 2 (0.1%) |
| Mania | 1 (<0.1%) |
| Uncategorized/other | 563 (18.54%) |
| **Condition and medical history** | |
| Family history of seizure | 359 (11.8%) |
| Brain surgery | 92 (3.0%) |
| Head trauma | 79 (2.6%) |
| Heart attack or stroke | 66 (2.2%) |
| Lung disease | 38 (1.3%) |
| Current pregnancy (of females) | 8 (<0.1%) |
| **Medication** | |
| Information not available/Unknown | 1,027 (33.8%) |
| No current medication | 897 (29.5%) |
| Anti-seizure medication (ASM) | 803 (26.4%) |
| Other | 100 (3.3%) |

| Psychiatric | 83 (2.7%) |
| --- | --- |
| Antibiotics | 33 (1.1%) |
| Combination regimens | 25 (<0.1%) |
| Cardiovascular | 17 (<0.1%) |
| **Findings (interpretable recordings)** | |
| Normal | 1,998 (68.5%) |
| Abnormal (Epileptiform) | 701 (24.0%) |
| Abnormal (Non-epileptiform) | 215 (7.4%) |
| Indeterminate | 1 (<0.1%) |

*Prevalence of Abnormal EEGs*

Among the 2,915 interpretable EEG recordings, 1,998 (68.5%) were classified as normal and 917 (30.2%) as abnormal. Of the abnormal recordings, 701 (76.4%) demonstrated epileptiform abnormalities and 215 (23.4%) demonstrated non-epileptiform abnormalities, including focal slowing, generalized slowing, and other nonspecific EEG changes (Figure 3). One recording had an indeterminate finding.

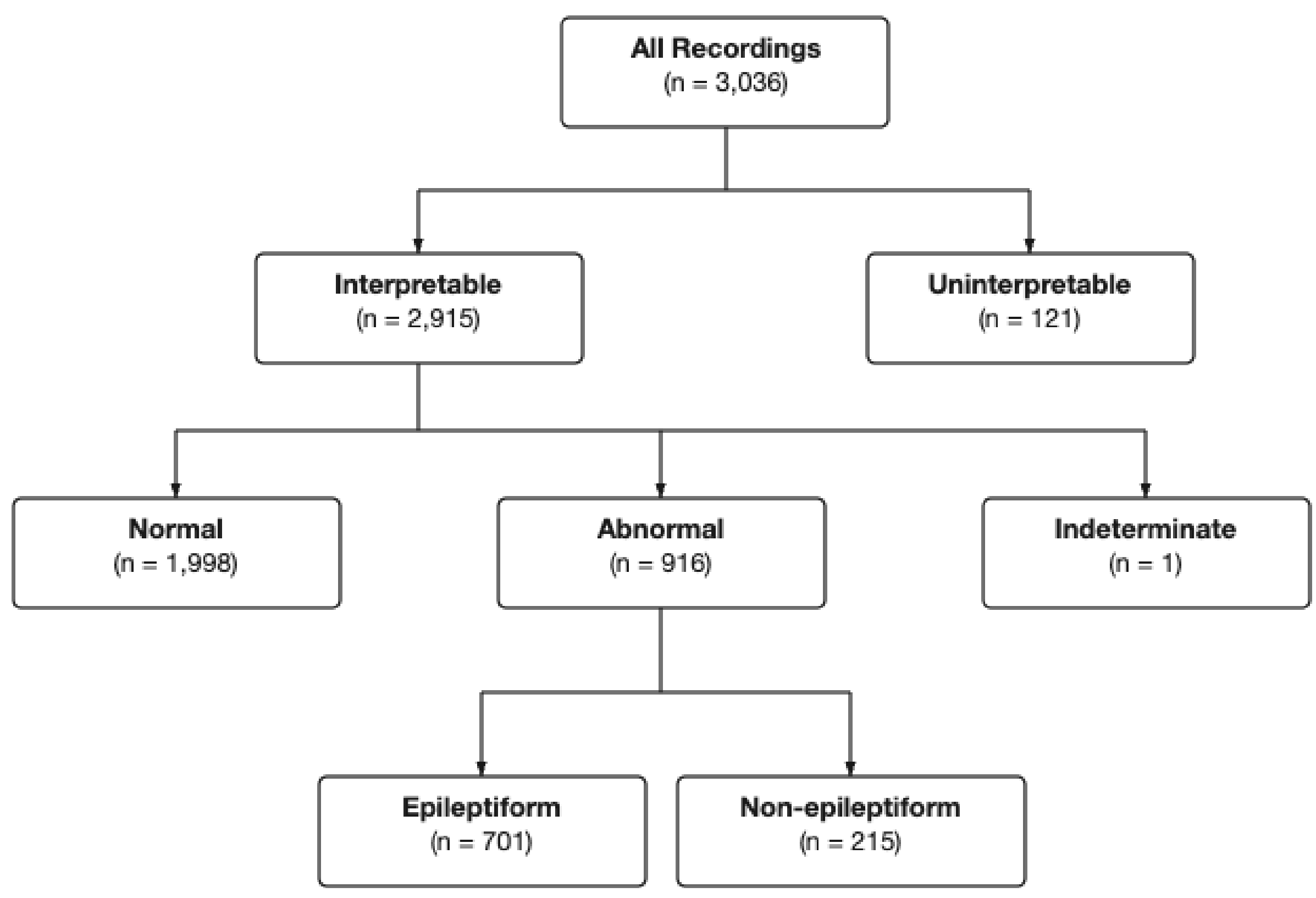

Figure 3. Flow diagram of EEG recordings and classification in the study cohort *(2-column)*

*Age-Specific Patterns of EEG Findings*

The distribution of EEG findings varied across age groups (Figure 4). Epileptiform abnormalities were most prevalent among children aged 4–9 years (n=170, 33.1%), followed by ages 10–19 (n=181, 25.6%) and 0–3 years (n=103, 21.9%) (Figure 4). Epileptiform rates were lower in adult age groups (14–21%). In contrast, non-epileptiform abnormalities were more common in participants aged ≥60 years (n=45, 19.2%) compared with younger age groups (3–9%).

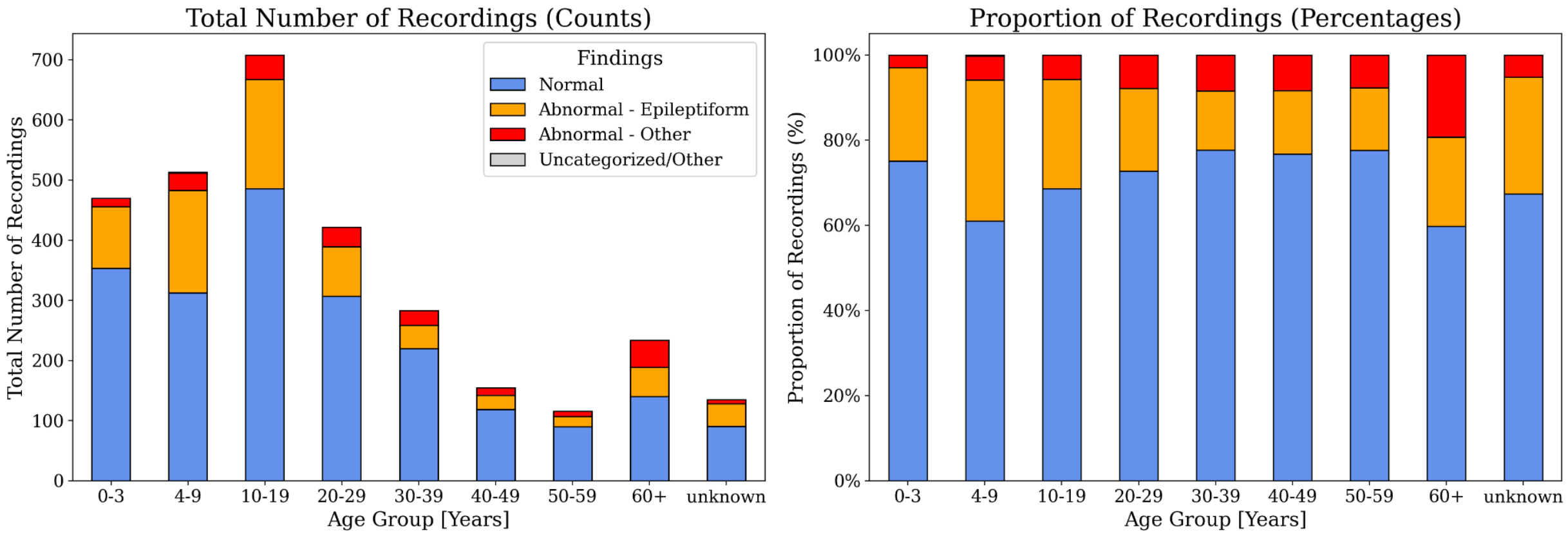

Figure 4. Age-Specific Distribution of Smartphone EEG Findings *(2-column)*

The distribution of abnormal findings varied across clinical indication categories. The most common indications for EEG referral were epilepsy-related presentations, particularly abnormal movements, seizures, or convulsions, which together accounted for 2,107 referrals (69.4% of all recordings) (Table 1). Behavioral and stroke-related referral indications demonstrated only non-epileptiform abnormalities in this dataset, while indications such as staring spells, sensory symptoms (e.g., paresthesias, numbness), and drop attacks were observed to include epileptiform abnormalities. Abnormal movements and other seizure-related problems represented the majority of clinical indications throughout all age groups.

Across most indication groups, epileptiform abnormalities represented the majority of abnormal findings. Patterns of abnormal findings by clinical indication are summarized in Figure 5.

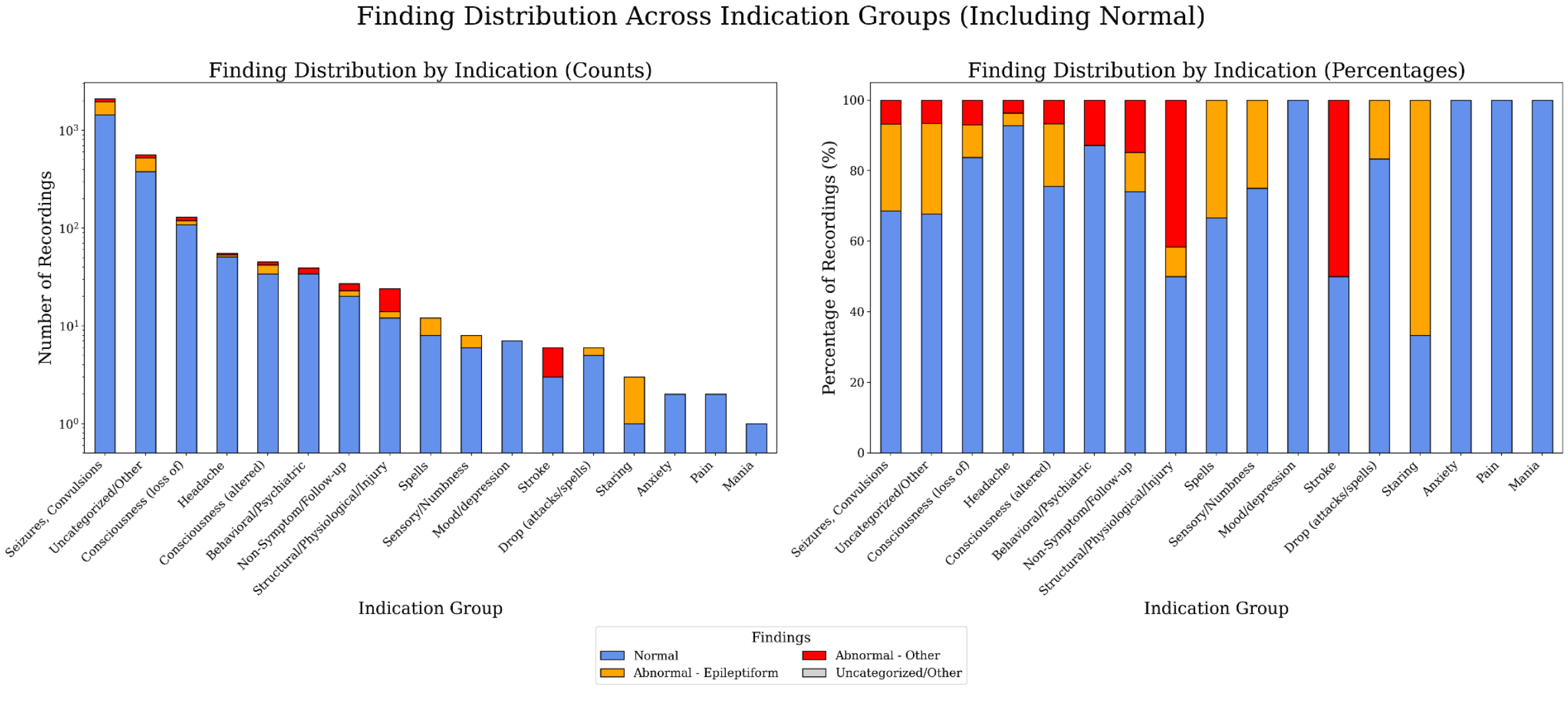

Figure 5. Abnormal Findings by Clinical Indication of the Study Cohort *(2-column)*

*Other Relevant Findings*
Among the 3,036 patients in the study, 999 (32.9%) had documented medication use at the time of their EEG recording. Of medicated patients, antiseizure medications were the most commonly reported class, with Phenobarbital (374), Valproate (142), Levetiracetam (92), and Carbamazepine (85) being the most frequently used agents. Medication status was unavailable or not recorded for the remaining 2,037 patients.

EEG recordings were distributed across 29 sites in Kenya, with highly uneven participation across sites. A single high-volume center contributed 1,658 recordings (55.0% of the Kenyan total), reflecting its role as the primary referral site in the network. The next four sites contributed a further 650 recordings (21.5%), and eight sites in total recorded 50 or more EEGs over the study period. The remaining 21 sites each contributed fewer than 50 recordings, with the smallest sites recording fewer than 10 EEGs. The median number of recordings per site was 31 (range 4–1,658), highlighting the degree of concentration.

**Discussion**

This study demonstrates findings from over 3,000 routine clinical smartphone-based EEG recordings performed across 29 sites in Kenya, representing one of the largest real-world evaluations of portable EEG implementation in a resource-limited health system. Several key findings emerged from this analysis. First, smartphone-based EEG technology displayed high feasibility in routine clinical care, with 96% of recordings being considered clinically interpretable by remote EEG interpreters. The device had high interpretability rates, rapid turnaround times for interpretation, and successful adoption across rural sites, highlighting scalability across resource-limited settings. The distribution of recordings was typical of early-adoption technology rollouts in low-resource settings, where initial uptake is driven by a small number of high-volume or highly motivated sites prior to broader scale-up. Second, abnormal EEG findings represented approximately a third of the interpretable recordings, with epileptiform abnormalities being found in 76.4% of all abnormal recordings. Third, age-related patterns showed that epileptiform abnormalities were observed more frequently among younger age groups, while older adults demonstrated a higher frequency of non-epileptiform abnormalities. Lastly, epilepsy-related clinical presentations represented the most common indication for EEG referral.

These findings suggest that the BC-1 smartphone EEG system has both clinical utility and high feasibility for implementation in decentralized clinical environments. The recordings provide meaningful diagnostic information that may support improved understanding of the epidemiology of epilepsy in Kenya and could further support broader patient populations in resource-limited settings.

The smartphone-based system used in this study was specifically designed to reduce technical complexity and simplify workflows in recording and interpretation. The prevalence of abnormal EEG findings observed in this study is consistent with patterns reported in prior studies of patients referred for seizure evaluation with epileptiform activity occurring in roughly 15-35% of routine EEGs and non-epileptiform abnormalities in 5-25%, suggesting that the diagnostic yield and clinical utility observed from smartphone-based EEGs fall within expected ranges [22–24]. Age-specific trends were consistent with existing data in global epidemiology. Epileptiform abnormalities were more frequently found among younger populations compared with adult age groups, reflecting the higher incidence of epilepsy and seizure disorders during childhood. In comparison, older adults demonstrated a greater frequency of non-epileptiform abnormalities, which could reflect age-related neurological conditions such as neurodegenerative disorders and systemic vascular changes [25]. Furthermore, the majority of clinical referral patterns in this sample were epilepsy-related presentations, which are consistent with typical clinical indications for EEG evaluation and suggest that smartphone-based EEG systems are being used in ways that parallel conventional EEG practice [26–27]. Lastly, many patients were either not receiving treatment or were not on ASMs, suggesting treatment gaps that access to smartphone EEG could help with.

The broader implications of these findings further global efforts to strengthen neurological care in resource-limited and underserved regions. Limited access to EEG infrastructure represents a key diagnostic barrier contributing to the inequities in seizure disorder underdiagnoses or misdiagnoses in LMICs. The high feasibility and diagnostic yield observed in this study provides practical evidence supporting the potential role of portable diagnostic technologies, aligning with the priorities outlined in the WHO's IGAP [3].

The study has several strengths. First, the multi-site nature of the study suggests that smartphone EEG helped extend clinical services beyond tertiary care centers into rural healthcare or even temporary facilities. Many of the 29 sites and imaging clinics included in the study had limited EEG capabilities previously, allowing recordings to be performed locally with interpretations occurring remotely. This model can help reduce transportation barriers, improve overall access to specialized diagnostic tools, and shorten diagnosis time, especially in countries where trained neurologists or healthcare resources are concentrated in urban areas. Second, the findings include over 3,000 clinical EEG recordings, representing one of the largest real-world datasets of smartphone-based EEG in a LMIC setting. Third, the data involves information on clinical indications, distributions of age, medication use, and more, allowing for a somewhat detailed observation of the patient population undergoing EEG evaluations across Kenya.

This study also involves several limitations. Firstly, the study was observational and descriptive in nature. It was not a population-based study. As smartphone EEGs were obtained from routine clinical referrals rather than systematically administered, participants were not chosen with a defined characteristic in mind, and clinical variables contained missing or incomplete data, including medication status and some demographic characteristics. Additionally, findings reflected patients who were selected for the smartphone EEG rather than the broader community of patients with prevalence of epilepsy. Secondly, the variability across the 29 participating sites was not monitored and accounted for. There was little data on the individual experience levels of the staff conducting the smartphone EEGs, so some healthcare workers could have more experience with screening and administering the recordings than others. Lastly, there were limited data on follow-up, preventing the assessment of how the smartphone EEGs influenced treatment. Additional research is needed to evaluate longer-term clinical outcomes and cost-effectiveness within health systems.

Future research should further explore the integration of smartphone EEG systems into national healthcare systems' infrastructure, potentially examining the clinical outcome, diagnosis timeline, and treatment initiation with having expanded EEG access. As well, implementation research examining training models and workflow integration of the system in diverse patient populations could provide valuable information for scaling across LMIC settings. Although no EEGs were interpreted using artificial intelligence, this remains an important future direction of this technology.

**Conclusion**

This study demonstrated that smartphone EEG can be successfully implemented within routine clinical practice across diverse clinical environments in Kenya. The findings provided a substantial number of interpretable EEGs showing meaningful detection of clinically relevant abnormal activities as well as the broad use across age groups and clinical indications suggesting a potential role in expanding neurodiagnostic access. By supporting the expansion of portable diagnostic systems, smartphone EEGs could potentially help reduce barriers to epilepsy diagnosis, improve treatment pathways, and inform future efforts to broaden equitable access to diagnostic EEG services in other resource-limited settings.

**Funding**

This research did not receive any specific grant from funding agencies in the public, commercial, or not-for-profit sectors.